\documentclass[preprintnumbers,amsmath,amssymb,12pt,floatfix,epsfig]{revtex4}
\usepackage{dcolumn}
\usepackage{bm}
\usepackage[dvips]{color}
\usepackage{graphicx}
\leftmargin -0.5in \baselineskip=24ptplus.5ptminus.2pt \vspace*{0.5
in} \large
\begin{document}
\title{Extended Optical Model Analyses of
Elastic Scattering and Fusion Cross Sections for
$^{6}$Li+$^{208}$Pb System at Near-Coulomb-Barrier Energies
by using Folding Potential}
\vspace{0.5cm}

\author{W. Y. So and T. Udagawa}
\address{Department of Physics, University of Texas,
Austin, Texas 78712}

\author{K. S. Kim}
\address{Department of Liberal Arts and Science, Hankuk
Aviation university, Koyang 412-791}

\author{S. W. Hong and B. T. Kim}
\affiliation{\it Department of Physics and Institute of Basic Science, \\
Sungkyunkwan University, Suwon 440-746, Korea}

\begin{abstract}
\begin{center}
{\vspace{1.0cm} \bf Abstract}
\end{center}
Based on the extended optical model approach in which the
polarization potential is decomposed into direct reaction (DR) and
fusion parts, simultaneous $\chi^{2}$ analyses are performed for
elastic scattering and fusion cross section data for the
$^{6}$Li+$^{208}$Pb system at near-Coulomb-barrier energies.
A folding potential is used as the bare potential. It is
found that the real part of the resultant DR part of the
polarization potential is repulsive, which is consistent with the
results from the Continuum Discretized Coupled Channel (CDCC)
calculations and the normalization factors needed for the folding
potentials.
Further, it is found that both DR and fusion parts of the
polarization potential satisfy separately the dispersion relation.
\end{abstract}
\maketitle \vspace{1.5cm} PACS numbers : 24.10.-i,~25.70.Jj

\pagebreak

\section{Introduction}

Much attention has been focused on two well known problems
originally revealed in the optical model analyses of the
elastic scattering data for loosely bound projectiles such as
$^{6}$Li and $^{9}$Be when a folding potential is used
for the real part of the optical potential~\cite{sat1,kee1}.
First, as demonstrated by Satchler and
Love~\cite{sat1}, it is needed to reduce the magnitude of the
folding potential by a factor $N=0.5 \sim 0.6$ to fit the data ; problem(1).
Secondly, the threshold anomaly~\cite{mah1,nag1} does not
appear in the resultant normalization constant $N$ fixed from the
fit to the data at near-Coulomb-barrier energies~\cite{kee1} ; problem(2).

It is natural to expect that these two problems may originate from
the strong breakup character of the loosely bound projectiles; in
fact, studies have been made of the effects of the breakup on the
elastic scattering, based on the coupled discretized continuum
channel (CDCC) method~\cite{sak1,kee2}. These studies were very
successful in reproducing the elastic scattering data without
introducing an arbitrary normalization factor and further in
understanding the physical origin of the factor $N=0.5 \sim 0.6$
needed when only one channel optical model calculations were made.
The authors of Refs.~\cite{sak1,kee2} projected their coupled
channel equations to a single elastic channel equation and deduced
the polarization potential arising from the coupling with the
breakup channels. The resultant real part of the polarization
potential was then found to be repulsive at the surface region
around the strong absorption radius, $R_{sa}$. This means that the
reduction of the folding potential by a factor of $N=0.5 \sim 0.6$
needed to be introduced when only one-channel optical model
calculation is made is to effectively take into account the effects
of the coupling with breakup channels. The CDCC studies, however,
have not been able to solve the problem (2) mentioned above, i.e.,
the fact that the normalization factor $N$ does not show the
threshold anomaly.

To solve the problem (2), it was suggested some time ago~\cite{uda1}
that the threshold anomaly is due to fusion: In case where fusion
is the dominant part of all the reaction processes, threshold anomaly
naturally manifests itself in the optical potential extracted from
the fit to elastic scattering data. However, in case where breakup
or direct reactions (DR) dominate, the energy dependence
of the resultant optical potential is governed by DR and thus should
be quite smooth~\cite{mah1}. In order to see the threshold anomaly
in the latter case, it is thus necessary to separate the polarization
potential into fusion and DR (breakup) parts. The threshold anomaly
will then be observed in the fusion part of the potential.

In order to test this idea, we have thus carried out~\cite{so1,kim1}
simultaneous $\chi^{2}$ analyses of elastic scattering and fusion
cross section data for the
$^{6}$He+$^{209}$Bi~\cite{agu1,kol1,agu2},
$^{6}$Li+$^{208}$Pb~\cite{kee1,wu1,das1}, and
$^{9}$Be+$^{209}$Bi~\cite{sig1,sig2} systems at near-Coulomb-barrier
energies in the framework of the extended optical
model~\cite{uda2,hong,uda3} that introduces two types of complex
polarization potentials, the DR and fusion potentials. In such
analyses, in addition to the elastic
scattering cross sections $d\sigma_{E}^{exp}/d\Omega$, the measured fusion
cross section $\sigma_{F}^{exp}$, was taken into account
together with the total experimental DR cross
section, $\sigma_{D}^{exp}$, if available, or the semi-experimental
DR cross section, $\sigma_{D}^{semi-exp}$, if $\sigma_{D}^{exp}$ was not available.

The DR and fusion potentials thus determined revealed some
characteristic features: First of all, both potentials satisfy
separately the dispersion relation~\cite{mah1}. Secondly, the fusion
potential is found to exhibit the threshold anomaly, as was observed
for tightly bound projectiles~\cite{bae1,lil1,ful1}, but the DR
potential does not show a rapid energy variation, i.e., the threshold
anomaly. Thirdly, at the strong absorption radius, the magnitude of
the fusion potential was found to be much smaller than that of the
DR potential. As a consequence, the resulting total polarization
potential dominated by the DR potential becomes rather smooth as a
function of the incident energy. This has explained the reason why
the threshold anomaly is not seen in the optical potentials
determined for systems involving loosely bound projectiles such as
$^{6}$He, $^{6}$Li, and $^{9}$Be~\cite{kee1,agu1,sig1}.

In the extended optical model analyses made so far~\cite{so1,kim1}
use was made of a rather shallow real potential for the bare
potential. The aim of the present study is to carry out for the
first time an extended optical model analysis of the elastic
scattering and fusion cross section data for the $^{6}$Li+$^{208}$Pb
system at near-Coulomb-barrier energies by utilizing a folding
potential as the bare potential. We shall show that the resulting
real part of the DR potential becomes repulsive and that the
threshold anomaly appears in the fusion potential, describing the
experimental data of the fusion and elastic scattering
cross sections without the two problems (1) and
(2) discussed in the beginning of the introduction.

In Sec. II, we first generate
$\sigma_{D}^{semi-exp}$ from the elastic scattering and fusion cross
section data by following the method proposed in Ref.~\cite{so1}.
$\chi^{2}$ analyses are then carried out in Sec. III, and the results
are presented and discussed in Sec. IV. Sec. V concludes the paper.

\section{Extracting semi-experimental DR cross section}

For the purpose of determining the fusion and DR
potentials separately, it is desirable to have data for the DR cross
section in addition to the fusion and elastic scattering cross
sections. For the $^{6}$Li+$^{208}$Pb system, however, no reliable
data for the DR cross section are available, although considerable
efforts have been devoted to measure the breakup and incomplete
fusion cross sections~\cite{sig3,sig4,sig5,liu1,das2}. We thus
generate the so-called semi-experimental DR cross section
$\sigma_{D}^{semi-exp}$, following the method proposed in
Ref.~\cite{so1}.

Our method to generate $\sigma_{D}^{semi-exp}$ resorts to the
empirical fact~\cite{kol1,mcc1,eli1,auc1,sin1} that the total
reaction cross section calculated from the optical model fit to the
available elastic scattering cross section data,
$d\sigma^{exp}_{E}/d\Omega$, usually agrees well with the
experimental $\sigma_{R}$, in spite of the well known ambiguities in
the optical potential. Let us call $\sigma_{R}$ generated
by the optical model the
semi-experimental reaction cross section $\sigma_{R}^{semi-exp}$.
Then, $\sigma_{D}^{semi-exp}$ is generated by
\begin{equation}
\sigma^{semi-exp}_{D} = \sigma^{semi-exp}_{R} - \sigma^{exp}_{F}.
\end{equation}
This approach seems to work even for loosely bound projectiles, as
demonstrated by Kolata {\it et al.}~\cite{kol1} for the
$^{6}$He+$^{209}$Bi system. We take $\sigma^{exp}_{F}$ from
Ref.~\cite{wu1}, but since the measured cross sections there
somewhat fluctuates as a function of energy, we smoothed out their
experimental cross sections using the Wong's formula~\cite{wong1}.

Following Ref.~\cite{so1}, we first carry out rather simple optical
model $\chi^{2}$ analyses of elastic scattering data solely for the
purpose of deducing $\sigma_{R}^{semi-exp}$. For these preliminary
analyses, we assume the optical potential to be a sum of $V_{0}(r)$
+ $i W_{I}(r)$ and $U_{1}(r,E)$, where $V_{0}(r)$ is the real,
energy independent bare folding potential to be discussed later in
III.B, $i W_{I}(r)$ is an energy independent short range imaginary
potential to be discussed in III.A, and $U_{1}(r,E)$ is a
Woods-Saxon type complex potential with common geometrical
parameters for both real and imaginary parts. The elastic scattering
data are then fitted with a fixed radius parameter $r_{1}$ for
$U_{1}(r,E)$, treating, however, three other parameters, the real
and the imaginary strengths $V_{1}$ and $W_{1}$ and the diffuseness
parameter $a_{1}$, as adjustable. The $\chi^{2}$ fitting is done for
three choices of the radius parameter; $r_{1}$=1.3, 1.4, and 1.5 fm.
These different choices of the $r_{1}$-value are made in order to
examine the dependence of the resulting $\sigma_{R}^{semi-exp}$ on
the value of $r_1$.

\begin{table}
\caption{Semi-experimental total reaction and DR cross sections for
the $^{6}$Li+$^{208}$Pb system.} \vspace{2ex} \label{semiexp}
\begin{ruledtabular}
\begin{tabular}{cccccc}
$E_{lab} $ &$E_{c.m.}$ & $\sigma^{exp}_{F}$ &
$\sigma^{semi-exp}_{D}$ & $\sigma^{semi-exp}_{R}$ &
$\sigma^{semi-exp}_{R}$ [2] \\ \hline
(MeV) & (MeV) & (mb) & (mb) & (mb) & (mb) \\
29 & 28.2 & 22 & 205 & 227 & 228 \\
31 & 30.1 & 120 & 306 & 426 & 431 \\
33 & 32.1 & 234 & 430 & 664 & 666 \\
35 & 34.0 & 335 & 545 & 880 & 897 \\
39 & 37.9 & 507 & 778 &1285 & 1303 \\
\end{tabular}
\end{ruledtabular}
\end{table}

As noted in Ref.~\cite{so1}, the values of $\sigma_{R}^{semi-exp}$
thus extracted for three different $r_{1}$-values agree with the
average value of $\sigma_R^{semi-exp}$ within 3\%, implying that
$\sigma_{R}^{semi-exp}$ is determined without much ambiguity. We
then identified the average values as the final values of
$\sigma_{R}^{semi-exp}$. Using thus determined
$\sigma_{R}^{semi-exp}$, we generated $\sigma_{D}^{semi-exp}$ by
employing Eq.~(1). The resultant values of $\sigma_{R}^{semi-exp}$
and $\sigma_{D}^{semi-exp}$ are presented in Table~\ref{semiexp},
together with $\sigma_{F}^{exp}$. In Table~\ref{semiexp}, given are
also $\sigma_{R}^{semi-exp}$ determined in Ref.~\cite{kee1}. 
The two sets of $\sigma_{R}^{semi-exp}$ determined
independently agree with each other. Note that in this study we
use the same normalization factors for the experimental
elastic scattering cross sections as in Ref.~\cite{kee1}. This
was not the case in Ref.~\cite{so1}, and thus the extracted
$\sigma_{D}^{semi-exp}$ in Table I and 
$\sigma_{D}^{semi-exp}$ in Ref.~\cite{so1} are
slightly different. In III.E, comparison will be made of
$\sigma_{D}^{semi-exp}$ thus extracted with the existing data for
breakup and incomplete fusion, and also the final calculated DR
cross section.

\section{Simultaneous $\chi^{2}$ Analyses}

Simultaneous $\chi^{2}-$analyses were then performed for the 
data sets of ($d\sigma^{exp}_{E}/d\Omega$,~$\sigma_{D}^{semi-exp}$,
~$\sigma^{exp}_{F}$) by taking 
$d\sigma^{exp}_{E}/d\Omega$, and $\sigma^{exp}_{F}$ from the
literature~\cite{kee1,wu1}. In calculating the $\chi^{2}$ value, we
simply assumed 1\% errors for all the experimental data. The 1\%
error is about the average of errors in the measured elastic
scattering cross sections, but much smaller than the errors in the
DR ($\sim$5\%) and fusion ($\sim$10\%) cross sections. The choice of
the 1\% error for DR and fusion cross sections is thus equivalent to
increasing the weight for the DR and fusion cross sections in
evaluating the $\chi^{2}$-values by factors of 25 and 100,
respectively. Such a choice of errors may be reasonable, since we
have only one datum point for each of these cross sections, while
there are more than 50 data points for the elastic scattering cross
sections.

\subsection{Necessary Formulae}

The optical potential $U(r,E)$ we use in the $\chi^{2}$-analyses has the
following form;
\begin{equation}
U(r;E) = V_{C}(r)-[V_{0}(r)+U_{F}(r;E)+U_{D}(r;E)],
\end{equation}
where $V_{C}(r)$ is the usual Coulomb potential with $r_{C}$=1.25 fm
and $V_{0}(r)$ is the bare nuclear potential, for which use is made
of the double folding potential to be described in more detail in
the next subsection. $U_{F}(r;E)$ and $U_{D}(r;E)$ are,
respectively, fusion and DR parts of the so-called polarization
potential~\cite{love} that originates from couplings to the
respective reaction channels. Both $U_{F}(r;E)$ and $U_{D}(r;E)$ are
complex and their forms are assumed to be of volume-type and
surface-derivative-type~\cite{kim1,hong}, respectively.
They are explicitly given by \begin{equation}
U_{F}(r;E) = (V_{F}(E)+iW_{F}(E))f(X_{F})+iW_{I}(r),
\end{equation}
and
\begin{equation}
U_{D}(r;E) = (V_{D}(E)+iW_{D}(E))4a_{D}\frac{df(X_{D})}{dR_{D}},
\vspace{2ex}
\end{equation}
where $f(X_{i})=[1+\mbox{exp}(X_{i})]^{-1}$ with
$X_{i}=(r-R_{i})/a_{i}$ $({\it i}=F\; \mbox{and} \; D)$ is the usual
Woods-Saxon function with the fixed geometrical parameters of
$r_{F}=1.40$~fm, $a_{F}=0.43$~fm, $r_{D}=1.47$~fm, and
$a_{D}=0.58$~fm, while $V_{F}(E)$, $V_{D}(E)$, $W_{F}(E)$, and
$W_{D}(E)$ are the energy-dependent strength parameters. Since we
assume the geometrical parameters of the real and imaginary
potentials to be the same, the strength parameters $V_{i}(E)$ and
$W_{i}(E)$ ($i=F$ or $D$) are related through a dispersion
relation~\cite{mah1},
\begin {equation}
V_{i}(E)=V_{i}(E_{s}) + \frac {E-E_{s}}{\pi } \mbox{P}
\int_{0}^{\infty} dE' \frac {W_{i}(E')}{(E'-E_{s})(E'-E)},
\vspace{2ex}
\end {equation}
where P stands for the principal value and $V_{i}(E_{s})$ is the
value of $V_{i}(E)$ at a reference energy $E=E_{s}$. Later, we will
use Eq.~(5) to generate the final real strength parameters
$V_{F}(E)$ and $V_{D}(E)$ using $W_{F}(E)$ and $W_{D}(E)$ fixed from the
$\chi^{2}$ analyses. Note that the breakup cross section may include
contributions from both Coulomb and nuclear interactions, which
implies that the direct reaction potential includes effects coming
from not only the nuclear interaction, but also from the Coulomb
interaction.

The second imaginary potential $W_{I}(r)$ in $U_{F}(r;E)$ given by
Eq.~(3) is a short-range imaginary potential of the Wood-Saxon type given by
\begin{equation}
W_{I}(r) = W_{I}f(X_{I}),
\end{equation}
with $W_{I}=40$~MeV, $r_{I}=1.0$~fm, and $a_{I}=0.30$~fm. This
imaginary potential is introduced in order to eliminate unphysical
survivals of lower partial waves at very small values of $r$ when
this $W_{I}(r)$ is not introduced. Because of the deep nature of the
folding potential $V_{0}$ used in this study and also because
$W_{F}(E)f(X_{F})$ energy-dependent imaginary part of $U_{F}(r;E)$
in Eq.~(3) turns out to be not strong enough, reflections of lower
partial waves appear which causes oscillations of 
$d\sigma_E / d\Omega$ at large angles, but physically
such oscillations should not occur. Thus $W_{I}(r)$ is introduced to
eliminate this unphysical reflection of lower partial waves. We may
introduce the corresponding real part $V_{I}f(X_{I})$, but we ignore
it here, simply because such a real potential does not affect physical
observables, which means that it is impossible to extract the
information of such a potential from analysing the experimental
data.

In the extended optical model, fusion and DR cross sections,
$\sigma_{F}$ and $\sigma_{D}$, respectively, are calculated by using
the following expression~\cite{uda2,hong,uda3,huss}
\begin {equation}
\sigma^{th}_{i} = \frac {2}{\hbar v} <\chi^{(+)}|
\mbox{Im}~[U_{i}(r;E)]|~\chi^{(+)}> \hspace{.5in}
(i=F\;\mbox{or}\;D),
\end{equation}
where $\chi^{(+)}$ is the usual distorted wave function that
satisfies the Schr\"{o}dinger equation with the full optical model
potential $U(r;E)$ in Eq.~(2). $\sigma^{th}_{F}$ and
$\sigma^{th}_{D}$ are thus calculated within the same framework as
$d\sigma_{E}/d\Omega$ is calculated. Such a unified description
enables us to evaluate all the different types of cross sections on
the same footing.

\subsection{The Folding Potential}

The double folding potential $V_{0}(r)$ we use
as the bare potential may be written as~\cite{sat1}
\begin{equation}
V_{0}(r)=\int d{\bf r}_{1} \int d{\bf r}_{2} \rho_{1}(r_{1})
\rho_{2}(r_{2}) v_{NN}(r_{12}=|\bf{r}-\bf{r}_{1}+\bf{r}_{2}|),
\end{equation}
where $\rho_{1}(r_{1})$ and $\rho_{2}(r_{2})$ are the nuclear matter
distributions for the target and projectile nuclei, respectively,
while $v_{NN}$ is the sum of the M3Y interaction that describes the
effective nucleon-nucleon interaction and the knockon exchange
effect given as
\begin{equation}
v_{NN}(r)=7999\frac{e^{-4r}}{4r}-2134\frac{e^{-2.5r}}{2.5r}-262
\delta (r).
\end{equation}
For $\rho_{1}(r)$ we use the following Woods-Saxon form taken from
Ref.~\cite{jag1}
\begin{equation}
\rho_{1}(r)=\rho_{0}/\left[1+\mbox{exp}\left(\frac{r-c}{z}\right)\right],
\end{equation}
with $c=6.624$~fm and $z=0.549$~fm, while for $\rho_{2}(r)$ the
following form is taken from Ref.~\cite{sat1};
\begin{equation}
\rho_{2}(r)=\frac{3}{8\pi^{3/2}} \left[\frac{1}{a^{3}}
\mbox{exp}(-\frac{r^{2}}{4a^{2}})
-\frac{c^{2}(6b^{2}-r^{2})}{4b^{7}}\mbox{exp}(-\frac{r^{2}}{4b^{2}})
\right],
\end{equation}
with $a=0.928$~fm, $b=1.26$~fm, and $c=0.48$~fm. The parameters for
the above $\rho_{1}(r)$ and $\rho_{2}(r)$ were fixed from the charge
density, but we assume they can be used for the matter density also. We then use
code DFPOT of Cook~\cite{coo1} for evaluating $V_{0} (r)$.

\subsection{Threshold Energies for Subbarrier Fusion and DR}

As in Ref.~\cite{so1}, we utilize as an important quantity the
so-called threshold energy $E_{0,F}$ and $E_{0,D}$ of subbarrier
fusion and DR, respectively, which are defined as zero intercepts of
the linear representation of the quantities $S_{i}(E)$, defined by
\begin{equation}
S_{i} \equiv \sqrt{E \sigma_{i}} \approx \alpha_{i} (E-E_{0,i})
\;\;\; (i=F \; \mbox{or} \; D),
\end{equation}
where $\alpha_{i}$ is a constant. $S_{i}$ with $i=F$, i.e., $S_{F}$
is the quantity introduced originally by Stelson {\it et
al.}~\cite{stel}, who showed that in the subbarrier region $S_{F}$
from the measured $\sigma_{F}$ can be represented very well by a
linear function of $E$ (linear systematics) as in Eq.~(12). In
Ref.~\cite{kim1}, we extended the linear systematics to DR cross
sections. In fact the DR data are also well represented by a linear
function.

\begin{figure}
\begin{center}
\vspace*{5.0cm}
\includegraphics[width=0.95\linewidth] {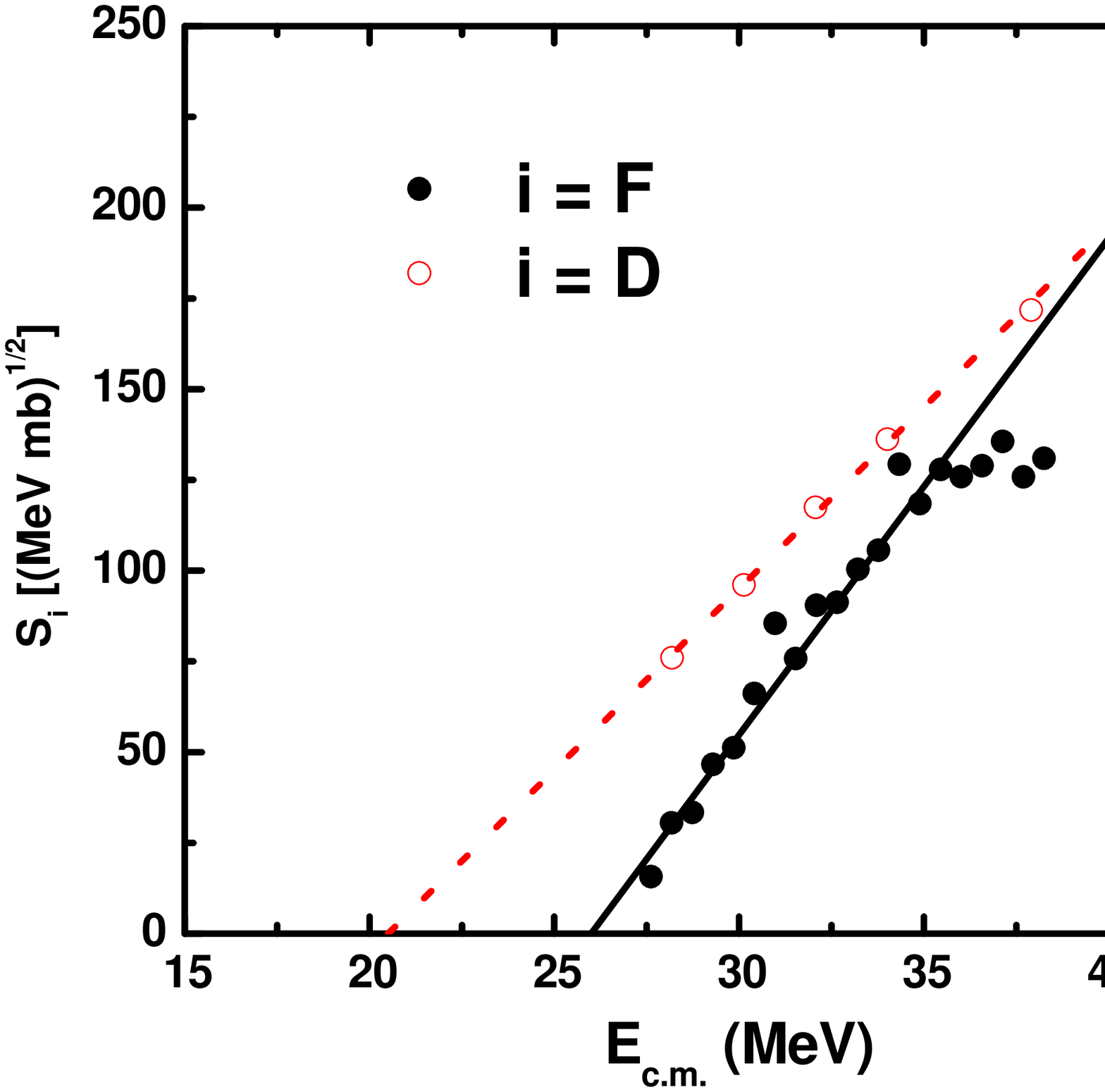}
\end{center}
\caption{\label{s-factor} The Stelson plot of
$S_{i}=\sqrt{E_{c.m.}\sigma_{i}}$ for DR ($i=D$, open circles) and
fusion ($i=F$, solid circles) cross sections. Use is made of the
semi-experimental DR cross section for $S_{D}$, while the
experimental fusion cross section is employed for $S_{F}$. The
intercepts of the straight lines allow us to extract the threshold
energies $E_{0,i}$.}
\end{figure}

In Fig.~\ref{s-factor}, we present the experimental $S_{F}(E)$ and
$S_{D}(E)$. For $S_{D}(E)$, use is made of $\sigma_{D}^{semi-exp}$.
For both $i=F$ and $D$, $S_{i}$ are very well approximated by
straight lines in the subbarrier region and thus $E_{0,i}$ can be
extracted without much ambiguity. 
From the zeros of $S_{i}(E)$, one can deduce
$E_{0,D}^{semi-exp}$=20.5~MeV and $E_{0,F}^{exp}=$26.0~MeV
in the c.m. system. 
It is interesting to note that
$E_{0,D}^{semi-exp}$ is found to be considerably smaller than
$E_{0,F}^{exp}$, meaning that the DR channels open at lower energies
than fusion channels, which seems physically reasonable.

$E_{0,i}$ may then be used as the energy where the imaginary
potential $W_{i}(E)$ becomes zero, i.e.,
$W_{i}(E_{0,i})=0$~\cite{kim1,kim2}. This procedure will be used
later in the next subsection for obtaining a mathematical expression
for $W_{i}(E)$.

\subsection{$\chi^{2}$ Analyses}

All the $\chi^{2}$ analyses performed in the present work are
carried out by using the folding potential as the bare potential
$V_{0}(r)$ described in III.B and by using 
the polarization potentials with fixed geometrical parameters, 
$r_{F}$=1.40~fm, $a_{F}$=0.43~fm, $r_{D}$=1.47~fm, and $a_{D}$=0.58~fm, 
which are close to the values used in our previous study~\cite{kim1}.
Some changes of the values from those of Ref.~\cite{kim1} were 
made to improve the $\chi^{2}$-fitting.

\begin{figure}
\begin{center}
\includegraphics[width=0.70\linewidth]{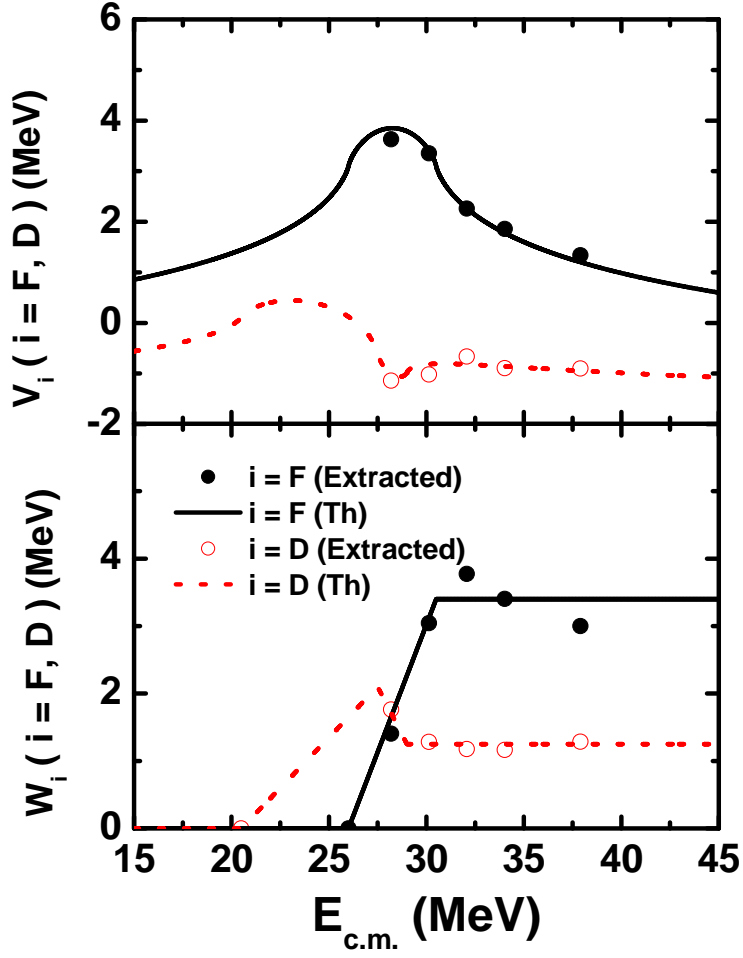}
\end{center}
\caption{\label{dispersion}~The strength parameters $V_{i}$ (upper
panel) and $W_{i}$ (lower panel) for $i=D$ and $F$ as functions of
$E_{c.m.}$. The open and the solid circles are the strength
parameters for $i=D$ and $F$, respectively. The dotted and the solid
lines in the lower panel denote $W_{D}$ and $W_{F}$ from Eqs. (13)
and (14), respectively, while the dotted and the solid curves in the
upper panel represent $V_{D}$ and $V_{F}$ calculated by using the
dispersion relation of Eq. (5).
The potential values and the corresponding reference energies
are such that $V_F$ ($E_s$ = 30.5 MeV) = 3.1 MeV and $V_D$ ($E_s$ =
27.5 MeV) = -0.6 MeV, respectively. }
\end{figure}

As in Ref.~\cite{kim1}, the $\chi^{2}$ analyses are done in two
steps; in the first step, all 4 strength parameters, $V_{F}(E)$,
$W_{F}(E)$, $V_{D}(E)$ and $W_{D}(E)$ are varied. In this first step, we
can fix fairly well the strength parameters of the DR
potential, $V_{D}(E)$ and $W_{D}(E)$, in the sense that $V_{D}(E)$
and $W_{D}(E)$ are determined as smooth functions of $E$. The
values of $V_{D}(E)$ and $W_{D}(E)$ thus extracted are presented in
Fig.~\ref{dispersion} by open circles. The values of $W_{D}(E)$ thus
extracted can be well represented by the following function of
$E(=E_{c.m.})$ (in units of MeV)
\begin{equation} W_{D}(E) \; = \; \left \{
\begin{array}{lll}
0 &\;\; \mbox{for $E\leq E_{0, D}^{semi-exp}=$20.5} \\
0.300(E-20.5) &\;\; \mbox{for 20.5$<E\leq$27.5} \\
-0.567(E-27.5)+2.10 &\;\; \mbox{for 27.5$<E\leq$29.0} \\
1.25 &\;\; \mbox{for 29.0$< E$} \\
\end{array}
\right. \vspace{2ex}
\end{equation}
Note that the threshold energies where $W_{D}(E)$ becomes zero are
set equal to $E_{0,D}^{semi-exp}$ as determined in the previous
subsection and are also indicated by the open circles at
$E=20.5$~MeV in Fig.~\ref{dispersion}. The dotted line in the lower
panel of Fig.~\ref{dispersion} represents Eq.~(13). The dotted curve in the
upper panel of Fig.~\ref{dispersion} denotes $V_{D}$ as predicted by
the dispersion relation of Eq.~(5), with $W_{D}(E)$ given by Eq.~(13).
As seen, the dotted curves reproduce the open circles fairly well,
indicating that $V_{D}(E)$ and $W_{D}(E)$ extracted by the
$\chi^{2}$ analyses satisfy the dispersion relation.

In this first step of $\chi^{2}$ fitting, however, the values of
$V_{F}(E)$ and $W_{F}(E)$ are not reliably fixed in the sense that
the extracted values fluctuate considerably as functions of $E$.
This is understandable from the expectation that the elastic
scattering data can probe most accurately the optical potential in
the peripheral region, which is nothing but the region characterized
by the DR potential. The part of the nuclear potential responsible
for fusion is thus difficult to pin down in this first step.

In order to obtain more reliable information on $V_{F}$ and $W_{F}$,
we thus have performed the second step of the $\chi^{2}$ analysis;
this time, instead of doing a 4-parameter search we fix $V_{D}$
and $W_{D}$ as determined by the first $\chi^{2}$ fitting, i.e.,
$W_{D}(E)$ given by Eq.~(13) and $V_{D}(E)$ predicted by the
dispersion relation. We then perform 2-parameter $\chi^{2}$
analyses, treating only $V_{F}(E)$ and $W_{F}(E)$ as adjustable
parameters. The values thus determined are presented in
Fig.~\ref{dispersion} by the solid circles. As seen, both $V_{F}(E)$ and
$W_{F}(E)$ are determined to be fairly smooth functions of $E$. The
$W_{F}(E)$ values may be represented by
\begin{equation}
W_{F}(E) \; = \; \left \{ \begin{array}{lll}
0 &\;\; \mbox{for $E\leq E_{0, F}^{exp}=$26.0} \\
0.756(E-26.0) &\;\; \mbox{for 26.0$<E\leq$30.5} \\
3.40 &\;\; \mbox{for 30.5$< E$} \\
\end{array}
\right. \vspace{2ex}
\end{equation}
As is done for $W_{D}(E)$, the threshold energy where $W_{F}(E)$
becomes zero is set equal to $E_{0,F}^{exp}$, which is also indicated
by the solid circle in Fig.~\ref{dispersion}. As seen, the
$W_{F}(E)$ values determined by the second $\chi^{2}$ analyses can
fairly well be represented by the functions given by Eq.(14). Note
that the energy variations seen in $W_{F}(E)$ and $V_{F}(E)$ are
more pronounced than those in $W_{D}(E)$ and $V_{D}(E)$ and
exhibit the threshold anomaly as observed in tightly bound
projectiles~\cite{bae1,lil1,ful1}.

Using $W_{F}(E)$ given by Eq.~(14), one can generate $V_{F}(E)$ from
the dispersion relation. The results are shown by the solid curve in
the upper panel of Fig.~\ref{dispersion}, which again well
reproduces the values extracted from the $\chi^{2}$-fitting. This
means that the fusion potential determined from the present analysis
also satisfies the dispersion relation.

\subsection{Final Calculated Cross Sections in Comparison with
the Data}

\begin{figure}
\begin{center}
\includegraphics[width=0.85\linewidth]{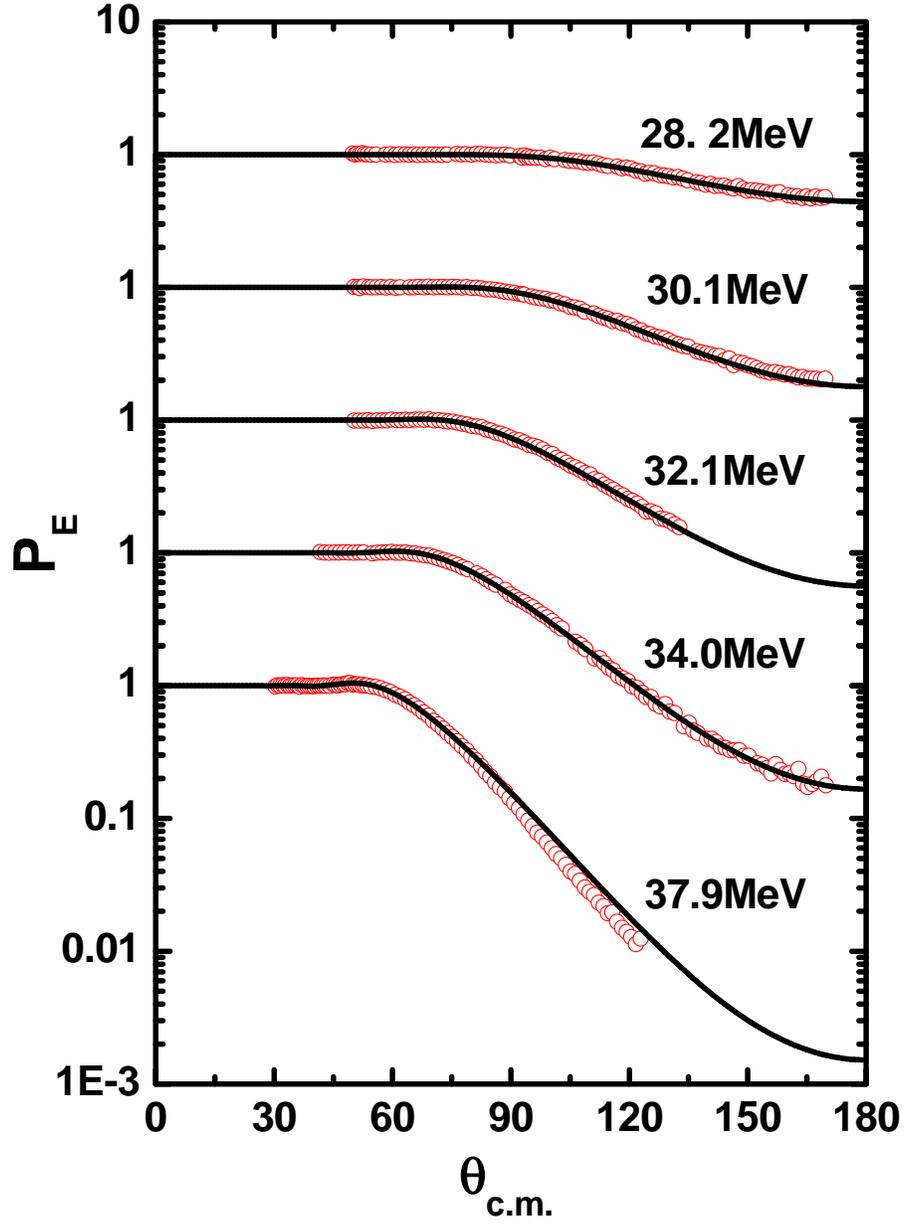}
\end{center}
\caption{\label{elastic}~Ratios of the elastic scattering cross
sections to the Rutherford cross section,
$P_{E}$=$\sigma_{E}/\sigma_{R}$, calculated with our final
dispersive optical potential are shown in comparison with the
experimental data. The data are taken from Ref.~\cite{kee1}.}
\end{figure}

\begin{figure}
\begin{center}
\vspace*{5.0cm}
\includegraphics[width=0.95\linewidth]{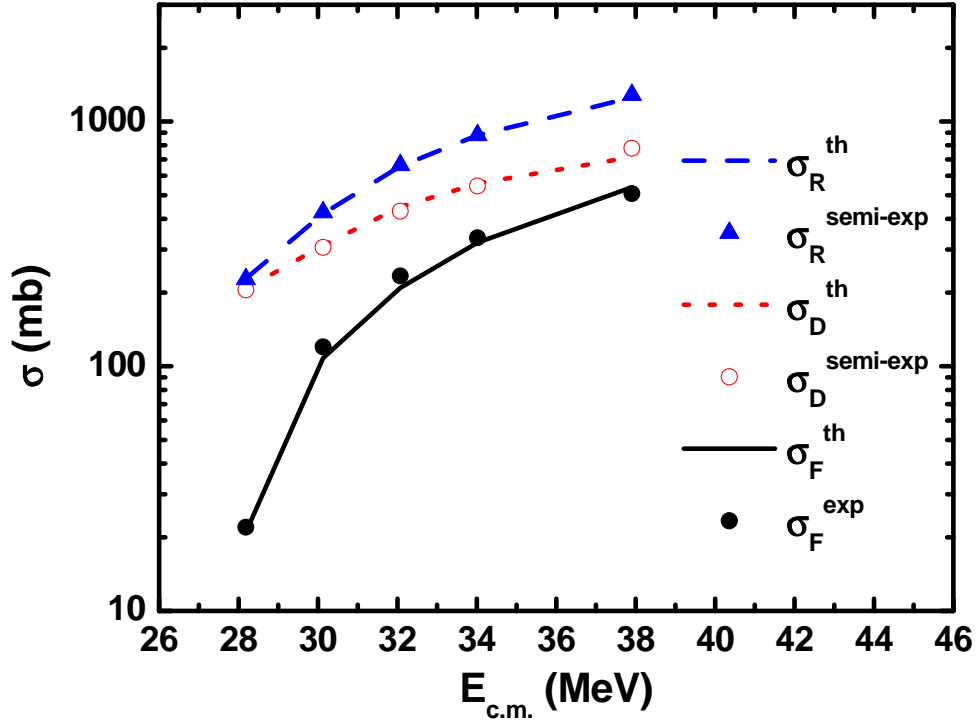}
\end{center}
\caption{\label{reaction} ~DR and fusion cross sections calculated
with our final dispersive optical potential are shown in comparison
with the experimental data. $\sigma_{R}^{semi-exp}$ and $\sigma_{D}^{semi-exp}$
denoted by the triangles and the open circles, respectively, are obtained as described in Sec.II. The fusion data
are from Ref.~\cite{wu1}.}
\end{figure}

Using $W_{D}(E)$ given by Eq.~(13) and $W_{F}(E)$ given by Eq.~(14)
together with $V_{D}(E)$ and $V_{F}(E)$ generated from the
dispersion relation, we have performed the final calculations of the
elastic, DR and fusion cross sections. The results are presented in
Figs.~\ref{elastic} and~\ref{reaction} in comparison with the
experimental data. All the data are well reproduced by the
calculations.

It may be worth noting here that the theoretical fusion cross
section, $\sigma_{F}^{th}$, includes partial contributions,
$\sigma_{I}$ and $\sigma_{F}$, from two imaginary components
$W_{I}(r)$ and $W_{F}(E)f(X_{F})$ in $U_{F}(r,E)$ given by Eq. (3).
In Table~\ref{fusion} the partial contribution from the $W_{I}(r)$
part, denoted by $\sigma_{I}$, are presented in comparison with the
total fusion cross section, $\sigma_{F}^{th}$. As seen, the
contribution from the inner part, $W_{I}$, amounts to $14 \sim
25$~\% of $\sigma_{F}^{th}$, which is relatively small but not
negligible. It should be remarked, however, that the real potential
$V_{I}(r)=V_{I}f(X_{I})$ corresponding to $W_{I}(r)$ does not
contribute at all to any cross section if the strength $V_{I}$ is
less than, say, 20~MeV. This justifies the fact that we have ignored
the $V_{I}(r)$ term.

\begin{table}
\caption{Partial contributions $\sigma_{I}$ and $\sigma_{F}$ in
comparison with the total fusion section, $\sigma^{th}_{F}$.}
\vspace{2ex} \label{fusion}
\begin{ruledtabular}
\begin{tabular}{ccccc}
$E_{lab} $ &$E_{c.m.}$ & $\sigma_{I}$ & $\sigma_{F}$ &
$\sigma^{th}_{F}$
\\ \hline
(MeV) & (MeV) & (mb) & (mb) & (mb) \\
29 & 28.2 & 5   & 16  &  21  \\
31 & 30.1 & 16  & 92  & 108  \\
33 & 32.1 & 29  & 180 & 209  \\
35 & 34.0 & 50  & 270 & 320  \\
39 & 37.9 & 100 & 439 & 539  \\
\end{tabular}
\end{ruledtabular}
\end{table}

At the moment, there are no data available for the DR cross
sections, $\sigma_{D}^{exp}$, which we may compare with our
calculated DR cross section $\sigma_{D}^{th}$ of Eq.~(7).
However, there are some data available; breakup-fusion cross sections (cross
sections of breakup of $^{6}$Li~$\longrightarrow \alpha + d$
followed by the absorption of one of the fragments) which is
referred to as the incomplete fusion cross section,
$\sigma_{ICF}$, in Ref.~\cite{liu1} and also exclusive $\alpha-d$ and
$\alpha - p$ coincidence cross sections~\cite{sig4}. The sum of
these cross sections become fairly large. In Table~\ref{direct}, we
present the sum of these cross sections denoted as
$\sigma_{ICF+excl}$ in comparison with our theoretical DR cross
sections. As seen, $\sigma_{ICF+excl}$ is slightly smaller than
$\sigma^{th}_{D}$, which is reasonable, since $\sigma_{ICF+excl}$
does not include such contributions as inelastic excitations of the
target nucleus and the incomplete fusion in which only a proton is
emitted, and so on. It is thus highly desirable to measure the remaining
missing parts of the DR cross sections in the future.

\begin{table}
\caption{Incomplete fusion plus exclusive coincidence cross
sections, $\sigma_{ICF+excl}$ in comparison with $\sigma_{D}^{th}$
for the $^{6}$Li+$^{208}$Pb system.} \vspace{2ex} \label{direct}
\begin{ruledtabular}
\begin{tabular}{ccccc}
$E_{lab} $ &$E_{c.m.}$ & $\sigma_{ICF+excl}$ & $\sigma^{th}_{D}$ & $\sigma^{semi-exp}_{D}$ \\
\hline
(MeV) & (MeV) & (mb) & (mb) & (mb) \\
29 & 28.2 & & 207 & 205 \\
31 & 30.1 & 264 & 312 & 306 \\
33 & 32.1 & 415 & 448 & 430 \\
35 & 34.0 & 517 & 558 & 545 \\
39 & 37.9 & 735 & 715 & 778 \\
\end{tabular}
\end{ruledtabular}
\end{table}

\subsection{Discussions}

\begin{figure}
\begin{center}
\vspace*{5.0cm}
\includegraphics[width=0.95\linewidth]{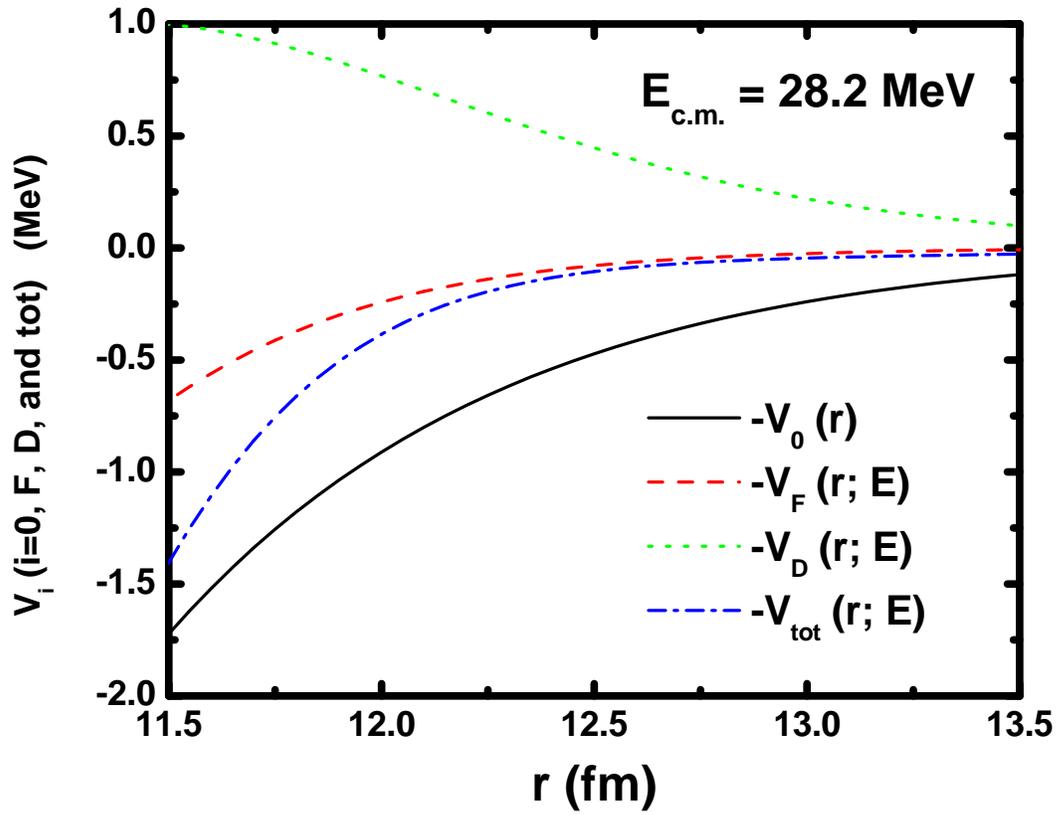}
\end{center}
\caption{\label{potential} The double folding potential, real parts
of fusion and DR potentials, and the sum of these potentials are plotted
for $E_{c.m.}$=28.2 MeV.}
\end{figure}

It is remarkable that the real part of the DR potentials determined
in the present $\chi^{2}$ analysis turn out to be repulsive at all
the energies considered here.
We present in Fig.~\ref{potential} the real part of the DR potential,
$-V_{D}(r,E)$, at $E_{c.m.}=28.2$~MeV in comparison with the folding
potential, $-V_{0}(r)$, in the surface region, $11.5$ fm $< r < 13.5$
fm. Also, the real part of the fusion
potential, $-V_{F}(r,E)$, and the sum, $-V_{tot}(r,E)$, of all these
three potentials are shown. As seen, the values of the sum of
real potentials are
significantly reduced from those of the bare folding potential.

\begin{figure}
\begin{center}
\vspace*{5.0cm}
\includegraphics[width=0.95\linewidth]{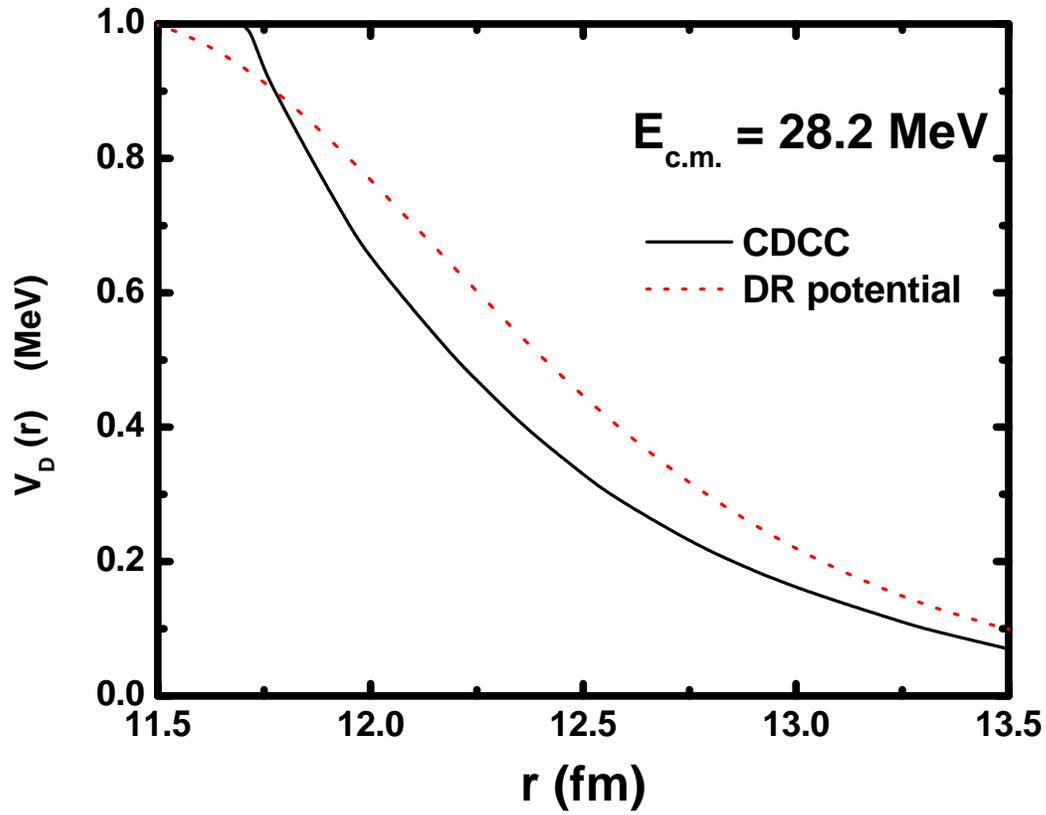}
\end{center}
\caption{\label{cdcc} The real part of the DR potential $V_{D}(r,E)$
is plotted
in comparison with that of the polarization potential deduced from the
CDCC calculations for $^{6}$Li+$^{208}$Pb system at $E_{c.m.}$=28.2
MeV.}
\end{figure}

It may be interesting to compare our $-V_{D}(r,E)$ with the real
part of the polarization potential obtained from the CDCC
calculations. Such a comparison is made in Fig.~\ref{cdcc}, where
$-V_{D}(r,E)$ shown in Fig.~\ref{potential} is compared with the
polarization potential calculated at $E_{c.m.}=28.2$~MeV \cite{kee2}.
As seen, two potentials show similar behaviours and
agree qualitatively with each other both in magnitude and in radial
dependence. This indicates that the DR potential deduced from the
present analyses of the elastic scattering and fusion data describes
essentially the same physical effects 
as treated in the CDCC calculation.

In Table~\ref{ratio} presented are the values of the strong
absorption radius $R_{sa}$, and those of $V_{0}$, $V_{F}$, $V_{D}$,
$V_{tot}$, $W_{F}$, $W_{D}$, and $R=V_{tot}/V_{0}$ at $r=R_{sa}$ for
all the energies considered here. The values of $R_{sa}$ decrease
slightly with the incident energy and range from 12.27~fm to
12.75~fm. Note that the value of $R= 0.19 \sim 0.37$ may be compared
with that of 0.51 obtained in Ref.~\cite{kee2}. (The normalization
factor used for the same system at $E_{lab}$ = 50.6 MeV in
Ref.~\cite{sat1} was 0.59.) It is seen also in Table~\ref{ratio}
that at the strong absorption radius $R_{sa}$, the values of the
real and the imaginary parts of the DR potential are both considerably
greater than those of the fusion potential. Because of this, the
energy dependence of the net polarization potential (sum of the
fusion and DR potentials) at $R_{sa}$ is dominated by that of the DR
potential with rather a smooth energy dependency. Consequently, the
net potential does not show such a threshold anomaly as seen in the net
potential for systems with tightly bound
projectiles~\cite{bae1,lil1,ful1}.

As already remarked, the real and the imaginary parts of both fusion and
DR potentials determined in the present $\chi^{2}$ analyses satisfy
well the dispersion relation, and further the fusion potential shows
clearly the threshold anomaly.

\begin{table}
\begin{center}
\caption{\small The value of the strong absorption radius, $R_{sa}$, and
those of $V_{0}$, $V_{F}$, $V_{D}$, $V_{tot}$, $W_{F}$, $W_{D}$, and
$R=V_{tot}/V_{0}$ evaluated 
at $r=R_{sa}$ for all the energies.} \vspace{2ex}
\label{ratio}
\begin{ruledtabular}
\begin{tabular}{c|c|c|c|c|c|c|c|c}
E$_{c.m.}$ & $R_{sa}$ & $-V_{0} $ & $-V_{F} $ & $-V_{D} $ & $-V_{tot} $ & $-W_{F} $ & $-W_{D} $& $R=V_{tot}/V_{0}$ \\
(MeV) & (fm) & (MeV) & (MeV) & (MeV) & (MeV) & (MeV) & (MeV)& \\
\hline
28.2 & 12.75 & -0.338 & -0.045 & 0.318 & -0.065 & -0.019 & -0.539 & 0.19 \\
30.1 & 12.56 & -0.436 & -0.061 & 0.334 & -0.163 & -0.056 & -0.512 & 0.37 \\
32.1 & 12.46 & -0.499 & -0.051 & 0.382 & -0.168 & -0.077 & -0.583 & 0.34 \\
34.0 & 12.39 & -0.547 & -0.047 & 0.437 & -0.157 & -0.090 & -0.635 & 0.29 \\
37.9 & 12.27 & -0.641 & -0.042 & 0.553 & -0.130 & -0.118 & -0.730 & 0.20 \\
\end{tabular}
\end{ruledtabular}
\end{center}
\end{table}

\section{Conclusions}

From the discussions of our results in the previous section, we may safely
conclude that within the extended optical model approach, even if
use is made of the double folding potential as its bare potential,
one can describe the elastic scattering and fusion cross section
data simultaneously without encountering the two problems remarked
at the beginning of this paper. The normalization factor needed to
be introduced to the folding potential,
particularly for loosely bound projectiles, in the earlier
analyses~\cite{sat1,kee1} based on the conventional optical model
approach can now be removed in the present extended optical model analysis,
and the effects are accounted for by means of
the repulsive DR potential as observed in the CDCC
approach. Also the threshold anomaly that could not be seen
in the analyses based on the conventional optical model approach is
now seen in the fusion part of the polarization potential.

In the present work, we focused our attention only to the
$^{6}$Li+$^{208}$Pb system, but it is possible to carry out similar
analyses to other systems. It is particularly interesting to do the analysis for
the $^{7}$Li+$^{208}$Pb system, where the conventional analysis is
successfully applied to explain the data. An extension of the
present analysis to the $^{7}$Li+$^{208}$Pb system is now under way,
and the report of the results will be made in a separated
paper.

This work was supported by the Korea Research Foundation Grant
funded by the Korean Government (MOEHRD) (KRF- 2006-214-C00014).

\newpage

\end{document}